# How can the use of different modes of survey data collection introduce bias? A simple introduction to mode effects using directed acyclic graphs (DAGs)

## Authors
Georgia D Tomova*[1], Richard J Silverwood[1], Peter WG Tennant[2,3], Liam Wright[1]

## Affiliations
1. Centre for Longitudinal Studies, University College London, UK
2. Leeds Institute for Data Analytics, University of Leeds, UK
3. School of Medicine, University of Leeds, UK

*Corresponding author:
Georgia D Tomova, UCL Centre for Longitudinal Studies, 55-59 Gordon Square, London, WC1H 0NU, United Kingdom; g.tomova@ucl.ac.uk

## Funding
This study was funded by the Survey Data Collection Methods Collaboration (known as Survey Futures), a research programme funded by the UK Economic & Social Research Council (ES/X014150/1). The UCL Centre for Longitudinal Studies is also supported by the UK Economic & Social Research Council (ES/W013142/1). The funder had no role in the study conceptualisation, writing, or decision to publish.

## Conflicts of interest
PWGT is a director of Causal Thinking Ltd which provides causal inference research consultancy and training; the company and the author may benefit from any work that demonstrates the value of causal inference methods. The other authors declare no conflicts of interest.

## Acknowledgements
None.

## Authorship (CRediT)
Conceptualisation: GDT, RJS, LW; Writing – original draft: GDT; Writing – review & editing: RJS, PWGT, LW; Funding acquisition – RJS, LW.

## Data availability statement
No data were used or generated as part of this manuscript.

**Abstract**

Survey data are self-reported data collected directly from respondents by a questionnaire or an interview and are commonly used in epidemiology. Such data are traditionally collected via a single mode (e.g. face-to-face interview alone), but use of mixed-mode designs (e.g. offering face-to-face interview or online survey) has become more common. This introduces two key challenges. First, individuals may respond differently to the same question depending on the mode; these differences due to measurement are known as 'mode effects'. Second, different individuals may participate via different modes; these differences in sample composition between modes are known as 'mode selection'. Where recognised, mode effects are often handled by straightforward approaches such as conditioning on survey mode. However, while reducing mode effects, this and other equivalent approaches may introduce collider bias in the presence of mode selection. The existence of mode effects and the consequences of naïve conditioning may be underappreciated in epidemiology. This paper offers a simple introduction to these challenges using directed acyclic graphs by exploring a range of possible data structures. We discuss the potential implications of using conditioning- or imputation-based approaches and outline the advantages of quantitative bias analyses for dealing with mode effects.

# Introduction

Survey data are a type of self-reported data collected directly from respondents via questionnaire or interview. Such data can be collected cross-sectionally at a single timepoint, at multiple timepoints from different groups of people ('repeat cross-sectional', e.g. US National Health and Nutrition Examination Survey[1]; UK National Diet and Nutrition Survey[2]), or longitudinally by repeatedly measuring the same participants over time (e.g. US Health and Retirement Study[3]; UK Millennium Cohort Study[4,5]). They differ from other types of data commonly utilised in epidemiology, such as routinely collected data from electronic health records, which do not involve direct collection from respondents. Survey data can be collected in a variety of ways, for example via a face-to-face, telephone, or video interview, or via a self-completed paper questionnaire or web survey. The means through which data are collected is referred to as the survey 'mode'. Traditionally, large surveys have employed a single preferred mode of data collection. However, recently, there has been an increased transition towards 'mixed-mode' (or 'multimode') data collection, where multiple modes are used[6,7]. This may happen concurrently, e.g. where participants are given the choice of participating by face-to-face interview or completing a web survey; or sequentially, e.g. where participants are initially offered a web survey and non-responders are followed up by a telephone survey. Longitudinal surveys may additionally involve a change in mode over different waves.

There are several potential benefits to employing mixed-mode data collection, including reduced costs, by shifting respondents to a cheaper mode, and increased participation rates, by allowing participants to respond via their preferred mode[7]. However, mixed-mode data collection may also introduce unintended consequences. In particular, it can introduce differences in the measurement of variables due to the different modes used. Systematic deviations in the observed values of a variable measured using different modes are a type of systematic measurement error (i.e. information bias[8]), commonly referred to as a 'mode effect'[9] (or 'mode measurement effect'[10]). Mode effects can arise for various reasons[11]. In the presence of an interviewer, respondents may be less willing to disclose sensitive information or may provide more socially desirable responses (so called 'interviewer effects'[12]). On the other hand, where no interviewer is present, respondents to self-completion questionnaires may be less thoughtful and provide less accurate responses to complete the questionnaire more quickly (so called 'satisficing'[13]). Differences in the responses across modes may also arise due to differences in how the questions and answers are presented (e.g., as text or verbally), and in what order they appear[14].

Although the predominant advice has focussed on preventing mode effects through appropriate questionnaire and survey design[11,15], several statistical approaches have also been proposed to reduce bias from mode effects *post hoc*. These include conditioning, imputation, and quantitative bias analysis[15–17]. However, the validity of each approach depends on additional assumptions about the underlying causal structure, including the absence (or sufficient control) of self-selection into mode. While mode effects are generally understood as a form of systematic measurement error, differences in responses between modes may also be related to the composition of participants in each mode, a situation known as 'mode selection' (or 'mode selection effects'). In other words, the survey mode might influence both *who* responds by each mode (the mode selection) and *how* they respond (the mode effect). Mode selection may occur for several reasons. Some respondents may not be eligible to respond by certain modes, for example, due to lack of internet access or phone number. In a concurrent mixed-mode survey, participants may self-select into a mode based on their own personal preference, for example younger respondents or those with more years of education are more likely to choose to respond by web[18]. In a sequential mixed-mode survey, selection into mode can also occur due to personal preference if the respondent is aware that their preferred mode will be offered later, or due to a delayed response, as early and late respondents, who respond by different modes, may differ in their characteristics[19]. Regardless of survey design, selection into mode therefore will almost universally be related to respondent characteristics.

The survey design, including the availability of modes offered and their order, can influence non-participation and therefore impact the external validity by changing the composition of the final sample available for analysis. However, the design can also impact the internal validity if the decision to participate via a particular mode is related to the exposure and outcome of interest, because mode would then be a collider on a path between the exposure and outcome. Conditioning on survey mode can therefore introduce collider bias[20]. The survey mode(s) hence potentially have implications for both internal and external validity[21].

The implications of not considering mode selection when attempting to deal with mode effects may be underappreciated. Directed acyclic graphs (DAGs) offer an intuitive way to depict and examine such methodological challenges, and have been used to unveil a variety of phenomena and paradoxes in epidemiology[20,22–24]. Some previous studies have considered mode effects and selection using simple causal graphs[25–27], albeit in a limited range of scenarios. In this article, we aim to clarify the challenges of handling mode effects and mode selection using DAGs by depicting a range of different scenarios in which survey mode may be related to exposure-outcome relationships of interest. Throughout, we discuss the potential implications of conditioning on mode and offer general recommendations on how best to deal with mode effects depending on the underlying structure.

## Depicting mode effects and mode selection using DAGs

Throughout, we use **X** and **X*** to differentiate between the latent (true) value of a variable (**X**) and the observed measure obtained by a survey (**X***). For example, **Fig. 1** depicts a latent variable **X** for which a measurement **X*** is obtained via a survey but measurement variation is introduced by the survey mode. In such instances, reducing this variation to obtain estimates as if only a single mode of data collection was used may be of interest. It is important to note that, even with a single mode, the measured value would still likely differ from the true value due to measurement error. However, for simplicity, we do not consider other sources of measurement error, which will also exist in practice. Throughout, we focus on scenarios where the interest is in estimating the causal effect of an exposure on an outcome.

Mode effects may be introduced in the measurement of any variable, but for simplicity, we will focus on a set of scenarios where mode effects apply to the exposure and/or outcome. Suppose we are interested in the extent to which smoking is affected by occupation. It may be plausible to assume that the reporting of occupation will be objective and therefore not differential by mode, but we would expect smoking to be subject to mode effects due to social desirability. In a DAG (**Fig. 2A**), mode is easily identifiable as a competing exposure, a variable that causes the outcome but not the exposure[28]. Competing exposures introduce error, but not bias, into the exposure-outcome relationship, and this error can be reduced by conditioning on mode (**Fig. 2B**). Scenarios where the opposite is true may also exist, i.e. the exposure may be subject to a mode effect, but not the outcome (**Fig. 2C**). The consequence of this is equivalent to regression dilution (or attenuation) bias, classically caused by random measurement error in the exposure[29,30]. In this case, however, the source of measurement variation is mode. Since the mode is observed, the regression dilution bias can be removed by conditioning on mode (**Fig. 2D**). This principle extends to settings where mode also introduces variation in one or more confounders. Measurement error in the confounders leads to imperfect conditioning and residual confounding[31]. However, conditioning on mode will resolve this. In many settings, both the exposure and the outcome may be subject to mode effects. When this is the case, it is easy to depict mode as a confounder of the relationship between the measured versions of the exposure and the outcome (**Fig. 2E**). Confounding by mode may be similarly removed by conditioning (**Fig. 2F**).

When mode effects occur, i.e. where mode introduces variation in the measurement of a variable, regardless of whether this refers to the exposure, the outcome, or both, the solution is straightforward: conditioning on mode will remove the variation that has been introduced. However, proceeding with conditioning on mode naïvely, without also considering mode selection, may actually increase bias. Consider a simplified scenario where the exposure and outcome are not subject to mode effects but are related to participation via a specific mode. For example, when estimating the effect of years in education on frequency of internet use, we may assume these variables are unlikely to be reported differentially across modes. However, both frequent internet users and those with higher education might be more likely to participate by web survey compared to others[32], meaning both the latent exposure and outcome cause mode selection (**Fig. 3A**). Since this turns mode into a collider, conditioning on mode would introduce collider bias and should be avoided[33] (**Fig. 3B**). However, if it is correctly recognised that conditioning on mode is not required in this scenario, due to the lack of any mode effects, then the presence of mode selection itself will not introduce bias.

It is reasonable to assume that in most scenarios there will be a degree of both mode selection and mode effects. This is because most variables of interest in the health and social sciences may also be related to willingness to participate via a given mode and reported differentially across modes due to their sensitive nature. For example, when studying the effect of depression on alcohol intake, both the exposure and the outcome might affect willingness to participate in a face-to-face interview rather than a self-completion questionnaire, while the presence of an interviewer is also more likely to induce more socially desirable responses[34]. In this scenario, mode will be simultaneously a collider for the latent exposure and outcome, and a confounder for their measured versions (**Fig. 3C**). While conditioning on mode will remove the confounding bias introduced by mode, it will also introduce collider bias (**Fig. 3D**). Whether the confounding bias or the collider bias will dominate will depend on the relative magnitude of the different biasing paths. However, the degree of collider bias is more likely to be strong when the collider is directly caused by the exposure and outcome[35,36], as in the example discussed here.

Even when mode is not a direct consequence of the exposure and the outcome, it is still important to consider the potential for collider bias because mode may still be a collider on a path between the exposure and outcome. For example, in the context of examining the effect of depression on diet, both variables are likely to be subject to mode effects, but it is unlikely that diet directly causes mode selection, meaning mode cannot be a direct collider for both the exposure and the outcome. However, mode will still be a collider for the exposure and any other causes of mode selection, such as socio-economic position (SEP), which may be unobserved (**Fig. 3E**). Without considering the wider context, it may not be obvious that conditioning on mode in this setting would still introduce collider bias if SEP also causes diet (**Fig. 3F**). This also applies in reverse, where only the latent outcome (but not the latent exposure) directly causes mode, or where mode is a mediator (or is strongly related to one)[21,37]. Other more complex structures such as M-bias also commonly exist in contexts where the characteristics of individuals are related to both their exposure and outcome, as well as their decision to participate[21]. M-bias scenarios also risk introducing collider bias, however, it has been argued that bias introduced by such structures tends to be smaller[35,38]. In practice, many of these longer collider paths that are opened by conditioning on mode may be closed by conditioning on relevant ancestor variables, including many classical confounders. However, the determinants of mode selection will not be known for all contexts. And even when known, and when rich data are available, accurately measuring and conditioning on all determinants of selection would also likely be extremely challenging.

Although the general principles can be straightforwardly extended to other settings, it is important to note that the examples considered so far are more readily compatible with cross-sectional surveys, where the exposure and outcome are measured at the same point in time. When using data from longitudinal surveys, it is common to select exposure and outcome variables collected at different points in time. This brings additional challenges. To examine this, we extend the example from Fig. 3C to a longitudinal setting with repeated measures of depression and alcohol intake at two time points. Suppose we are interested in the total causal effect of the exposure at time 1 (***Depression$_1$***) on the outcome at time 2 (***Alcohol intake$_2$***). There may be two general scenarios that may occur with respect to mode. Firstly, the repeated measures of mode itself may be correlated due to participants being more likely to respond using a mode they have previously used (**Fig. 4A**). This will introduce shared variation in the measurements of the exposure and the outcome over time and bias their apparent association. Secondly, where changes in the availability of different modes occur over time, switching from one mode to another may mean that variables are measured differently at each time point. This can induce an apparent change over time that, although attributable to the mode switch, may be naively misinterpreted as meaningful. For example, the effect of ***Depression$_1$*** on ***Alcohol intake$_2$*** may be biased towards the null (**Fig. 4B**). To resolve this, researchers might consider conditioning on the different modes at the different time points. However, as in the previous examples, this is likely to introduce collider bias (**Fig. 4C**). Similarly to the example in Fig. 3F, this may in theory be resolved by conditioning on relevant ancestor variables. However, determinants of mode selection may again be unknown or unmeasured, necessitating the need for other statistical approaches to be considered as potential solutions.

## Discussion

### Key messages

The mode of survey data collection can introduce two distinct problems in the analyses of survey data: systematic measurement error (so called 'mode effects') and selection bias (a potential consequence of 'mode selection'). Using DAGs, we have shown that it is intuitive to understand the implications of these, wherever plausible assumptions can be made regarding the underlying causal structure. We show that whenever survey mode introduces measurement variation in one or more variables of interest, but is otherwise unaffected by selection, the consequences can be reduced by conditioning on mode. However, whenever differential selection according to mode occurs – which may be the norm rather than the exception – then conditioning on mode as an attempt to reduce mode effects would risk introducing collider bias.

**An overview of statistical solutions**

It is generally recommended that mode effects should be minimised prospectively through optimal questionnaire design in mixed-mode surveys[11], although it has also been argued that aiming to harmonise different modes may reduce their individual strengths[15]. Nevertheless, some differences are inherent and intractable – such as differences in whether items are presented verbally or as written text. The presence of an interviewer may even influence responses in self-completion modules[39]. As a result, a number of statistical solutions have been considered to address these challenges in the analytical stages[15–17,40]. We outline the general types of approaches below.

*Conditioning-based approaches*

Conditioning approaches (e.g. covariate adjustment or stratification) are widely used and understood in epidemiology. The benefits, and unintended consequences, of conditioning on survey mode are discussed in detail in this article. As described, the main challenge of conditioning on mode is the presence of mode selection. Where all back-door paths can be closed by conditioning on causes of mode selection, then researchers can proceed with conditioning on mode to remove measurement bias. However, this may be difficult to achieve in practice as it requires knowing about and measuring all these causes. It is also not a solution when the latent outcome directly causes mode selection because such bias cannot be resolved only by conditioning on common causes. Conditioning on mode should therefore be avoided when it is caused by the latent outcome unless the strength of the selection arc is deemed negligible. This is likely to be a particular concern for cross-sectional surveys where a latent outcome may substantially precede the response and might therefore be more likely to affect selection into mode. The principles of conditioning extend to other mathematically equivalent approaches, such as two-stage approaches of regressing a variable on mode and using the mode-adjusted value in subsequent analyses, which also carry the additional caveat that the standard errors must be correctly propagated.

*Imputation-based approaches*

Multiple imputation has also been suggested as a potential statistical solution[16,41]. This approach involves identifying the reference mode (i.e. the one considered more accurate or used historically in previous waves of a longitudinal or repeated cross-sectional survey) and imputing values for the alternate mode using information about the relationships between relevant variables from those who responded in the reference mode. Although multiple imputation is a routine approach for dealing with missing data, the data in this context are not missing, they are simply observed under different modes. In practice, this means that data from the alternate mode are discarded and replaced with imputed values, which is arguably wasteful. More important, however, is the assumption that data are missing at random. This would be violated under a 'missing not at random' (MNAR) scenario, which would occur whenever the variable being imputed directly causes mode selection, or whenever the common causes of mode selection and the variable are unobserved or unconditioned, since discarding the values from one mode implicitly conditions on mode. This is equivalent to the assumption of no mode selection required for conditioning-based approaches.

*Quantitative bias analysis approaches*

Alternatively, external information can be used to reduce or better understand the likely impact of survey mode on the estimates of applied studies. The impact of measurement bias can be reduced using standard calibration techniques for addressing measurement error[42]. In an ideal scenario, direct calibration information would be available from a validation study in a sub-sample (e.g. where a sub-group of participants respond by both modes), or from an external validation study in a comparable population. The approach would involve specifying a calibration equation in the validation sub-sample, which is then used to calibrate measures in the full sample (see *Boe et al.* 2023[43]). In practice, such validation studies are rare, though re-interview data has been used in a similar way[44,45]. However, a form of calibration can be attempted from other sources of external information. For example, if an estimate is available on the size of a mode effect and its uncertainty (e.g. from a study where mode is randomised), then this can be used to inform a simulation to produce bias-corrected estimates, assuming the mode effect estimate is transportable to the study sample. When calibration is not achievable, researchers can, alternatively, evaluate the likely impact of a hypothesised mode effect using simple sensitivity analysis techniques or by simulating scenarios under different possible mode effect sizes and evaluating the impact that this would have on the study results[46,47]. Both available or hypothesised mode effect estimates can be subject to effect modification and therefore vary across population sub-groups. Where an implausibly large mode effect would be required to alter the study results, then this provides reassurance that the findings are not sensitive to mode effects. Despite the requirement for some external information on the likely size of the mode effect, the main benefit of the calibration approach is that it does not require any assumptions regarding mode selection.

**Implications for applied research**

The implications for measurement and collider bias of using different modes of data collection are well understood in the survey methodology literature[9,48–50], although they are rarely if ever discussed using causal diagrams, meaning the full implications of conditioning on mode under different contextual scenarios may still be underappreciated. Although relatively straightforward and overlapping with existing principles, the existence and implications of mode effects have not been discussed extensively in the epidemiological literature, except for some studies estimating the magnitude of mode effects[51,52]. This is an oversight as epidemiological research is often conducted using data from large surveys, and increasingly these are employing mixed modes of data collection. Being unaware of the challenges this brings, or being aware but using statistical solutions without considering the assumptions, risks introducing bias in the estimated relationships. Since many surveys have implemented mixed-mode data collection in one or more sweeps, bias may have already appeared in the existing applied literature. This paper therefore offers a simple introduction to these issues using DAGs, which are intuitive to epidemiologists, with the aim of increasing awareness of this challenge and encouraging good practices.

Whether the mode of survey data collection will introduce measurement variation and/or will be subject to selection is context specific. There exist theoretical frameworks that can guide researchers towards understanding the risk of mode effects[11,53]. No equivalent frameworks exist for mode selection, however many of the existing recommendations on the use of DAGs already encourage researchers to explicitly consider selection nodes[54]. In general, it is plausible to assume that, to some extent, both mode effects and mode selection will exist. What matters, in practice, is the extent of bias that these issues are likely to introduce. If the mode effect is likely to be trivial, then it may be reasonable to ignore the use of different modes, regardless of any mode selection. Alternatively, if the likely mode selection is small, then conditioning on mode may be a reasonable strategy to remove non-trivial mode effects. If, however, meaningful mode effects and mode selection biases are likely, then quantitative bias analyses may be the only reasonable approach. It is hence important for researchers to learn to recognise when meaningful mode effects are likely to occur. Mode effects are particularly likely in mixed-mode settings where one mode is interviewer-led and the other is self-completed. This is because, in the presence of an interviewer, respondents are more likely to provide socially desirable answers and less likely to disclose private or otherwise sensitive information, for example when being asked about their mental health, personal relationships or behaviours[11]. Responses may also depend on the presentation of questions and answers, including whether options are read out by the interviewer or the participant, their specific order, and the type of categories or response scales offered. For example, if answer options are read out over the telephone, respondents are more likely to focus on and select one of the final options[55]. Whenever questions may be perceived as burdensome or requiring cognitive effort (e.g. open-ended questions, long or repetitive questions and answers, or questions requiring significant thought or recall), respondents may not give considered and thoughtful responses, especially in a self-completion questionnaire[56].

**Summary and recommendations**

In the analyses of survey data, it is important to consider the potential implications stemming from the use of different modes of data collection. Researchers are encouraged to consider the extent to which both measurement and collider bias may be present and communicate any assumptions regarding the hypothesised causal structure in their specific context of interest. DAGs offer an intuitive approach to do so, as well as to understanding the implications of different potential solutions. When mode selection is thought to be negligible or where the causes of mode selection can also be conditioned on, measurement bias may be reduced by conditioning on survey mode. However, researchers must proceed with caution whenever mode selection is relevant to the exposure-outcome relationship of interest. Conditioning on mode in these settings may introduce collider bias. Both conditioning- and imputation-based approaches require strong assumptions regarding selection that may not be reasonable in many situations. Quantitative bias analyses approaches should therefore be considered more widely to calibrate against, or understand, the potential impact of measurement variation introduced by survey mode.

# Figures

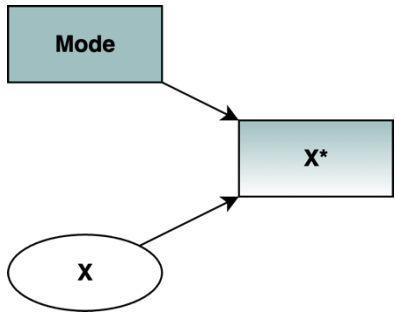

**Figure 1. A directed acyclic graph depicting measurement error introduced by survey mode.** A measure (***X*****) of a latent variable (***X***) is obtained via a survey but variation is introduced by the survey mode (***Mode***).

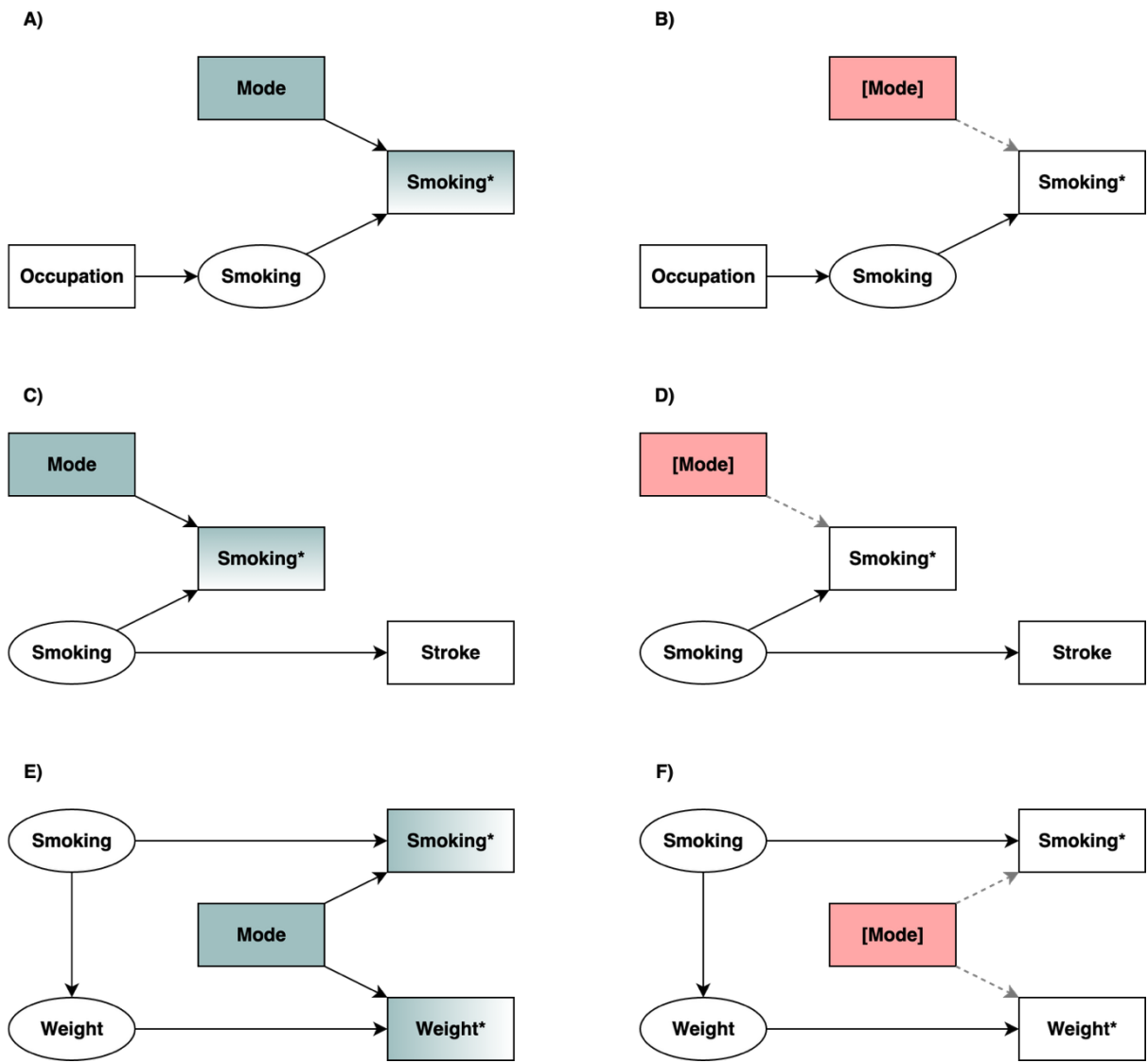

**Figure 2. A directed acyclic graph depicting survey mode effects on the exposure, the outcome, or both.** Shared variation is depicted in green; conditioned nodes are depicted in red and with square brackets. **A)** considers occupation and smoking as example exposure and outcome, where ***Smoking*** is a measured version of the latent ***Smoking*** with additional variation introduced by ***Mode***, leading to heterogeneity in the effect estimate. **B)** shows conditioning on ***Mode*** will reduce this heterogeneity by closing the path ***Mode -> Smoking****. **C)** considers smoking and stroke as example exposure and outcome, where ***Smoking**** is a measured version of the latent ***Smoking*** with additional variation introduced by ***Mode,*** leading to regression dilution in the effect estimate. **D)** shows conditioning on ***Mode*** will reduce the regression dilution by closing the path ***Mode -> Smoking****. **E)** considers smoking and body weight as example exposure and outcome, where ***Smoking**** and ***Weight**** are measured versions of the latent ***Smoking*** and ***Weight*** with additional common variation introduced by ***Mode***, creating confounding between ***Smoking**** and ***Weight****. **F)** shows conditioning on ***Mode*** will reduce the confounding by closing the path ***Smoking* <- Mode -> Weight*.***

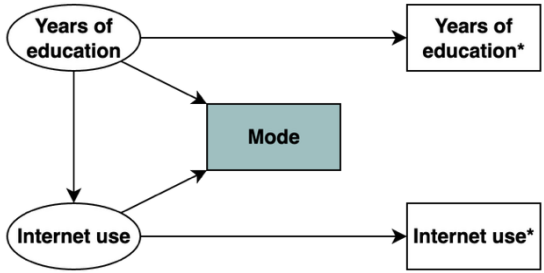
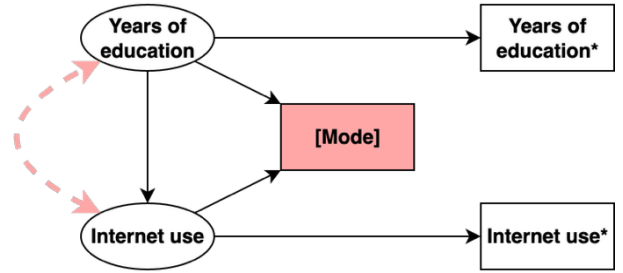
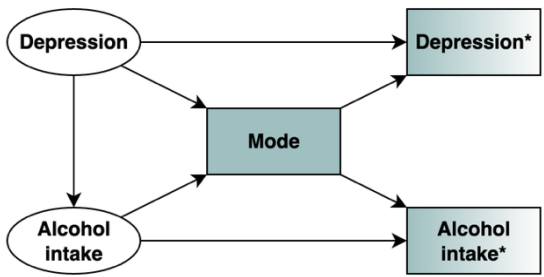
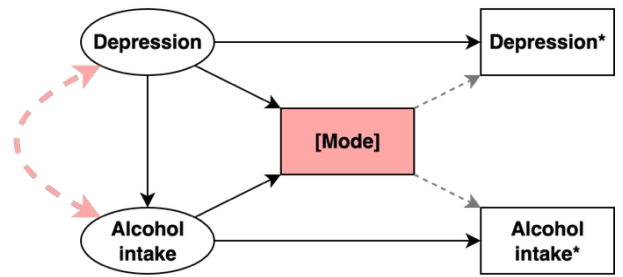
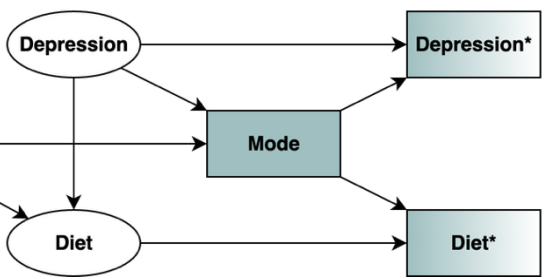
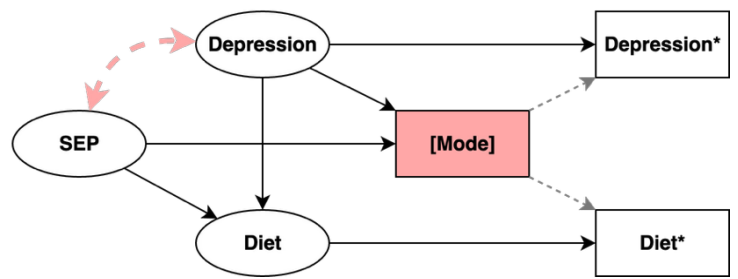

**Figure 3. A directed acyclic graph depicting mode selection with and without mode effects.** Shared variation is depicted in green; conditioned nodes are depicted in red and with square brackets. **A)** considers years of education and internet use as example exposure and outcome, where *Years of education\** and *Internet use\** are measured versions of the latent *Years of education* and *Internet use*, which themselves cause *Mode* selection. **B)** shows the consequences of naively conditioning on *Mode*, which will open the path *Years of education -> Mode <- Internet use*, introducing collider bias. **C)** considers depression and alcohol intake as example exposure and outcome, where *Depression\** and *Alcohol intake\** are measured versions of the latent *Depression* and *Alcohol Intake* with additional shared variation introduced by *Mode*, which creates confounding between *Depression\** and *Alcohol intake\**. The latent *Depression* and *Alcohol intake* in turn cause *Mode* selection. **D)** shows that conditioning on *Mode* will reduce confounding by closing the path *Depression\* <- Mode -> Alcohol intake\** but will introduce collider bias by opening the path *Depression -> Mode <- Alcohol intake*. **E)** considers depression and diet as example exposure and outcome, where *Depression\** and *Diet\** are measured versions of *Depression* and *Diet* with additional variation introduced by *Mode*, which creates confounding between *Depression\** and *Diet\**. *Mode* is caused by the latent *Depression*. The unobserved socio-economic position (*SEP*) is a common cause of *Mode* and *Diet*. **F)** shows that conditioning on *Mode* will reduce confounding by closing the path *Depression\* <- Mode -> Diet\** but will introduce collider bias by opening the path *Depression -> Mode <- SEP -> Diet*.

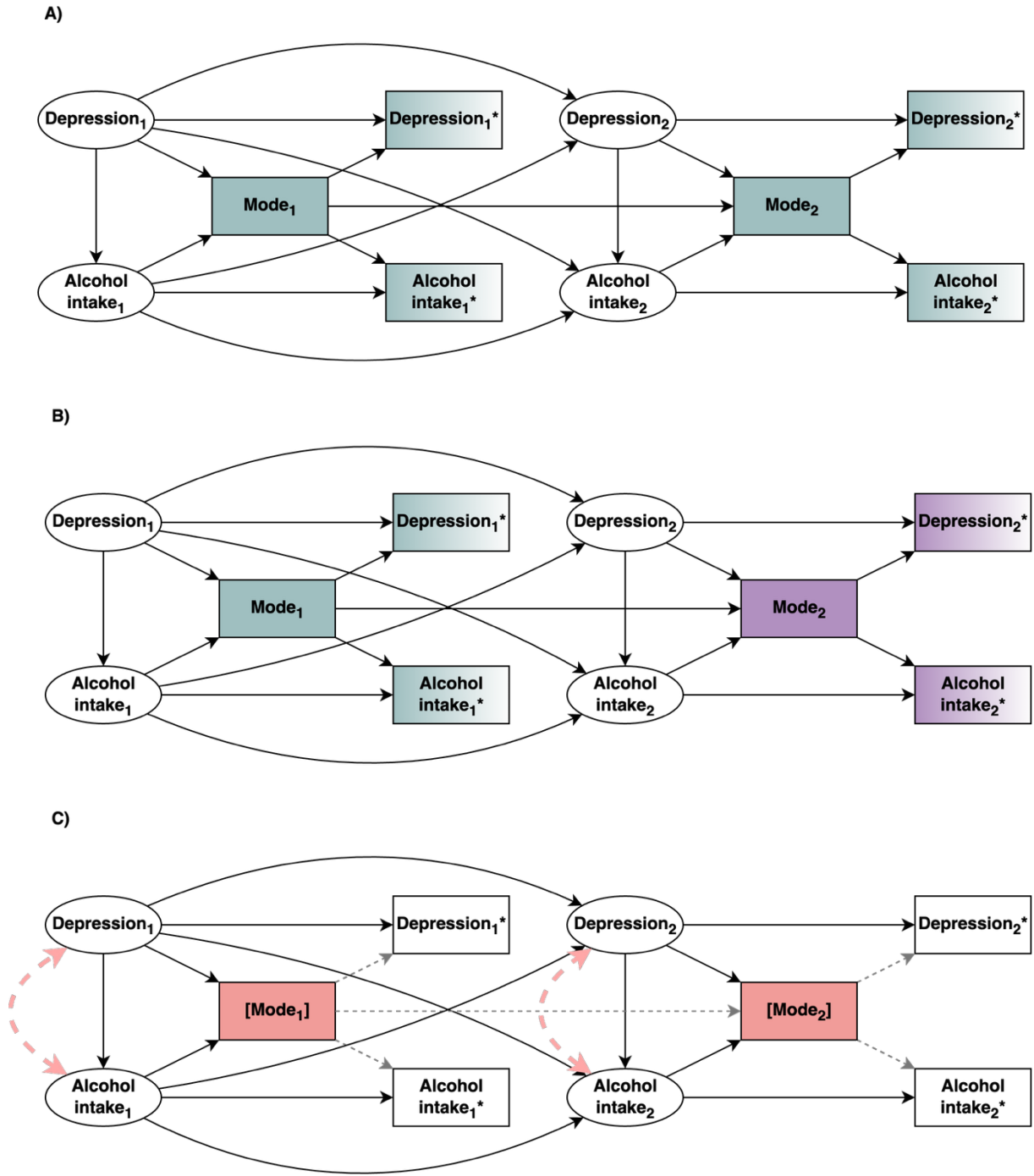

**Figure 4. A directed acyclic graph depicting longitudinal scenarios where both mode effects and mode selection occur.** Shared variation is depicted in green; conditioned nodes are depicted in red and with square brackets. **A), B) and C)** consider depression and alcohol intake where **$Depression_1*$**, **$Alcohol\ intake_1*$**, **$Depression_2*$**, **$Alcohol\ intake_2*$** represent measured versions of the latent **$Depression_1$, $Alcohol\ intake_1$, $Depression_2$, $Alcohol\ intake_2$** at two time points, and where **$Depression_1$** and **$Alcohol\ intake_2$** are the example exposure and outcome, respectively. Common variation in their measurement is introduced by the survey mode at both time points (**$Mode_1$** and **$Mode_2$**), which creates confounding between depression and alcohol intake at each time point. **A)** shows that when the survey modes are the same at both time points, common variation will be introduced (in green), biasing the effect estimate. **B)** shows that when the survey modes are different between the two time points, different sources of variation will be introduced at each time point (in green and purple), biasing the effect estimate towards the null. **C)** shows that, if **$Mode_1$** and/or **$Mode_2$** are conditioned, collider bias will be introduced by opening the path(s) **$Depression_1$ -> $Mode_1$ <- $Alcohol\ intake_1$** and/or **$Depression_2$ -> $Mode_2$ <- $Alcohol\ intake_2$**.